\documentclass{article}

\usepackage{arxiv}

\usepackage[utf8]{inputenc}        
\usepackage[T1]{fontenc}           

\usepackage{amsmath, amssymb, amsfonts, amsthm}  
\usepackage{nicefrac}             
\usepackage{siunitx}              

\usepackage{booktabs}             
\usepackage{multirow}             
\usepackage{array, tabularx}      
\usepackage{adjustbox}            

\usepackage{graphicx}             
\usepackage{float}                
\usepackage{subfigure}            
\usepackage{sidecap}              
\usepackage{newfloat}             
\DeclareFloatingEnvironment[name={Supplementary Figure}]{suppfigure}

\usepackage{tikz}
\usepackage{pgfplots}
\pgfplotsset{compat=1.18}

\usepackage{microtype}            
\usepackage{lipsum}               
\usepackage[numbers,square]{natbib}   
\usepackage{doi}                     
\usepackage{url}

\usepackage{xcolor}                 
\usepackage{lineno}                

\title{When Deep Learning Fails: Limitations of Recurrent Models on Stroke-Based Handwriting for Alzheimer’s Disease Detection}

\date{}

\newif\ifuniqueAffiliation
\uniqueAffiliationtrue

\usepackage{authblk}

\newbox{\orcid}\sbox{\orcid}{\includegraphics[scale=0.06]{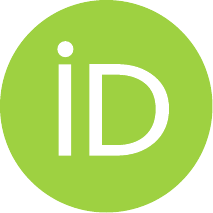}} 
\author[1]{%
	\href{https://orcid.org/0009-0005-8718-5435}{\usebox{\orcid}\hspace{1mm}Emanuele Nardone\thanks{\texttt{emanuele.nardone@unicas.it}}}%
}
\author[1]{%
	\href{https://orcid.org/0009-0005-4340-7227}{\usebox{\orcid}\hspace{1mm}Tiziana D'Alessandro\thanks{\texttt{tiziana.dalessandro@unicas.it}}}%
}
\author[1]{%
	\href{https://orcid.org/0000-0002-3242-0179}{\usebox{\orcid}\hspace{1mm}Francesco Fontanella\thanks{\texttt{fontanellao@unicas.it}}}%
}
\author[1]{%
	\href{https://orcid.org/0000-0002-7654-6849}{\usebox{\orcid}\hspace{1mm}Claudio De Stefano\thanks{\texttt{destefano@unicas.it}}}%
}
\affil[1]{Department of Electrical and Information Engineering (DIEI), University of Cassino and Southern Lazio, Cassino, Italy}

\renewcommand{\shorttitle}{When Deep Learning Fails on Stroke-Based AD Detection}

\begin{document}
\maketitle

\begin{abstract}
	Alzheimer's disease detection requires expensive neuroimaging or invasive procedures, limiting accessibility. This study explores whether deep learning can enable non-invasive Alzheimer's disease detection through handwriting analysis. Using a dataset of 34 distinct handwriting tasks collected from healthy controls and Alzheimer's disease patients, we evaluate and compare three recurrent neural architectures (LSTM, GRU, RNN) against traditional machine learning models. A crucial distinction of our approach is that the recurrent models process pre-extracted features from discrete strokes, not raw temporal signals. This violates the assumption of a continuous temporal flow that recurrent networks are designed to capture. Results reveal that they exhibit poor specificity and high variance. Traditional ensemble methods significantly outperform all deep architectures, achieving higher accuracy with balanced metrics. This demonstrates that recurrent architectures, designed for continuous temporal sequences, fail when applied to feature vectors extracted from ambiguously segmented strokes. Despite their complexity, deep learning models cannot overcome the fundamental disconnect between their architectural assumptions and the discrete, feature-based nature of stroke-level handwriting data. Although performance is limited, the study highlights several critical issues in data representation and model compatibility, pointing to valuable directions for future research.
\end{abstract}

\keywords{Deep Learning \and Handwriting analysis \and Alzheimer's disease}

\section{Introduction}
Deep learning (DL) approaches have revolutionised the analysis of sequential and temporal data across numerous domains \cite{lecun2015deep}. Recurrent neural architectures, including Recurrent Neural Networks (RNNs), Long Short-Term Memory (LSTM) networks, and Gated Recurrent Units (GRUs), are specifically engineered to capture time-varying patterns by processing information across continuous input sequences. These models are highly effective in applications where transitions and temporal relationships hold critical semantic or physiological significance, such as natural language processing and biomedical signal analysis.

However, the effectiveness of such architectures relies on the assumption that input sequences maintain temporal continuity and reflect underlying dynamic processes. This assumption is violated when data consists of discrete, pre-segmented events, such as handwriting strokes' feature vectors. The resulting temporal fragmentation can harm the ability of recurrent models to capture the dynamics they were designed to model, potentially compromising their performance in critical applications.

This study investigates this architectural mismatch in a clinically relevant context: detecting Alzheimer's Disease (AD) from digitised handwriting samples. This application presents an important test case, as handwriting intrinsically involves continuous motor control processes.

AD affects over 55 million people worldwide and represents the most common form of dementia, characterised by progressive cognitive decline and neurodegeneration \cite{alzheimer2021facts}. Early detection is crucial for timely intervention and treatment planning. However, current diagnostic methods, including neuroimaging techniques, cerebrospinal fluid analysis, and comprehensive cognitive assessments, remain expensive and inaccessible in many healthcare settings \cite{jack2018nia}. This accessibility gap underscores the urgent need for alternative, cost-effective screening methods.

Recent advances in digital biomarker research have identified handwriting analysis as a promising avenue for early AD detection \cite{thomas2017digital}. Handwriting requires complex neuromotor coordination that integrates cognitive processing, motor planning, and executive functions, all showing early compromise in AD progression \cite{yan2008handwriting}. Modern tablet-based acquisition systems enable precise capture of temporal dynamics, pressure variations, and kinematic features that may reveal subtle neurological changes imperceptible to human observation. These digital traces encode rich information about cognitive status within the fine-grained structure of handwriting movements.

Despite the theoretical promise of applying DL to handwriting-based AD detection, the practical implementation faces a fundamental challenge: handwriting data is typically processed as discrete strokes rather than continuous trajectories. While computationally convenient, this pre-segmentation may discard critical temporal information that recurrent architectures require for optimal performance.

This paper evaluates whether standard recurrent architectures can effectively detect AD from stroke-level handwriting features and diagnose the architectural limitations that emerge from this discrete representation. Specifically, we implement and compare three fundamental recurrent architectures (LSTM, GRU, and RNN) using a robust experimental framework that includes comprehensive hyperparameter optimisation and subject-level cross-validation to ensure generalizability across diverse patient populations. Our results reveal that direct application of these models to stroke-level features yields suboptimal performance, highlighting a critical mismatch between the discrete nature of the preprocessed data and the continuous temporal modelling assumptions of recurrent networks.

Through this analysis, we identify key methodological challenges in the field, including the absence of standardised stroke definitions and the loss of temporal coherence during segmentation. We propose future research directions that use raw time-series modelling and improved segmentation strategies to better preserve the temporal dynamics essential for accurate AD detection. Our findings have broader implications for applying DL to biomedical time-series data, particularly when clinical constraints need discrete event representations of inherently continuous processes.

The paper is organised as follows. Section \ref{sec:related} reviews related work on handwriting analysis for Alzheimer's disease detection and the use of deep learning in this domain. Section \ref{sec:materials} describes the dataset and the handwriting tasks considered. Section \ref{sec:methods} details our methodology, including feature extraction, model architectures, and evaluation strategy. Section \ref{sec:results} presents and discusses the experimental results. Finally, Section \ref{sec:conclusions} concludes the paper and outlines future research directions.

\section{Related Work}
\label{sec:related}

The concept of digital biomarkers has gained attention in neurodegenerative disease research, as they provide objective, quantifiable measures of biological processes through digital health technologies \cite{goldsack2020verification}. Handwriting analysis stands out as a promising method due to its non-invasive nature and the detailed temporal information it offers about motor and cognitive function. Prior work has examined various handwriting changes in patients with AD. For instance, Impedovo et al. \cite{impedovo2019handwriting} showed that kinematic features from handwriting tasks can distinguish AD patients from healthy controls. In a similar vein, Garre-Olmo et al. \cite{garre2017kinematic} found that certain temporal patterns in writing tasks align with cognitive assessment scores.

Traditional machine learning methods for detecting AD through handwriting have focused on feature engineering and standard classifiers. Drotár et al. \cite{drotar2016decision} applied decision trees and support vector machines to hand-crafted kinematic features, yielding moderate classification results. Yet these methods often demand domain knowledge for feature selection and may overlook complex temporal patterns in handwriting data. More recent work has sought to address these limitations by analysing individual handwriting strokes rather than aggregated features. In~\cite{nardone2025handwriting}, authors extracted dynamic and static features from on-paper and in-air movements across 34 tasks, using ensemble methods such as ranking-based and stacking approaches. Their stroke-level analysis achieved up to 80.18\% accuracy in AD detection and outperformed aggregated statistical methods on most tasks.

DL has produced strong results when applied to sequential biomedical data in various fields. Recurrent neural networks, such as LSTM and GRU models, have worked well for time-series analysis in healthcare \cite{choi2016retain}. Recent developments in attention mechanisms and Transformer models have improved the handling of long-range dependencies and the detection of key temporal patterns \cite{vaswani2017attention}.

\section{Materials}
\label{sec:materials}
In this study, we utilised a dataset of handwriting samples to study potential biomarkers for AD through temporal dynamics analysis. The materials contained the participant data and the specialised equipment for data capture.

We analysed handwriting data from 174 participants, including 89 individuals diagnosed with AD and 85 healthy controls (HC). This setup allowed us to frame the analysis as a binary classification task. To reduce the influence of confounding factors, we matched the groups closely on key demographic variables such as age, gender, education level, and occupation type. These characteristics are summarised in Table \ref{tab:demographics}. Participants with AD were recruited from the Alzheimer's Disease Unit at Federico II hospital in Naples, where they underwent standard clinical evaluations using tools like the Mini-Mental State Examination (MMSE) \cite{folstein1975mmse}, the Frontal Assessment Battery (FAB) \cite{dubois2000fab}, and the Montreal Cognitive Assessment (MoCA) \cite{nasreddine2005moca}. These assessments helped ensure that only those with cognitive impairments linked to AD were included, excluding cases with unrelated cognitive issues.

\begin{table}[t]
\centering
\caption{Demographic characteristics of study participants.}
\label{tab:demographics}
\renewcommand{\arraystretch}{1.1}
\begin{tabular}{@{}lcccc@{}}
\toprule
Group & Age & Education & Women & Men \\ 
\midrule
AD patients & 71.5 (9.5) & 10.8 (5.1) & 46 & 44 \\
HC & 68.9 (12.0) & 12.9 (4.4) & 51 & 39 \\
\bottomrule
\end{tabular}
\end{table}

We applied strict exclusion criteria to maintain the quality of the dataset. Individuals on psychotropic medications or other drugs that could affect cognitive function were not included. Medical specialists also screened for conditions that might alter handwriting, such as arthritis or Parkinson's disease, to keep the focus on features specific to AD.

The handwriting tasks consisted of 25 main activities divided into three categories, as outlined in Table \ref{tab:handwriting_tasks}. Graphic tasks tested basic stroke production through exercises like connecting dots and tracing shapes. Copy and reverse copy tasks involved replicating more intricate patterns, such as letters, words, and numbers. Memory and dictation tasks explored how the writing changed when participants recalled information or transcribed from visual cues.

\begin{table}[t]
\centering
\caption{List of the tasks. Memory and dictation (M), Graphic (G), and Copy (C).}
\label{tab:handwriting_tasks}
\adjustbox{width=\columnwidth}{
\setlength{\tabcolsep}{4pt}
\renewcommand{\arraystretch}{1.1}
\begin{tabular}{@{}clc@{}}
\toprule
\textbf{Task \#} & \textbf{Description} & \textbf{Category} \\
\midrule
1 & Signature drawing & M \\
2 & Join two points with a horizontal line, continuously four times & G \\
3 & Join two points with a vertical line, continuously four times & G \\
4 & Retrace a circle (6 cm in diameter) continuously four times & G \\
5 & Retrace a circle (3 cm in diameter) continuously four times & G \\
6 & Copy the letters 'l', 'm', and 'p' & C \\
7 & Copy the letters on the adjacent rows & C \\
8 & Write a sequence of four lowercase letters 'l' cursive & C \\
9 & Write a sequence of four lowercase cursive bigram 'le' cursively & C \\
10 & Copy the word "foglio" & C \\
11 & Copy the word "foglio" above a line & C \\
12 & Copy the word "mamma" & C \\
13 & Copy the word "mamma" above a line & C \\
14 & Memorise the words "telefono", "cane", and "negozio" and rewrite them & M \\
15 & Copy in reverse the word "bottiglia" & C \\
16 & Copy in reverse the word "casa" & C \\
17 & Copy six words (regular, non-regular, non-words) in the appropriate boxes & C \\
18 & Write the name of the object shown in a picture (a chair) & M \\
19 & Copy the fields of a postal order & C \\
20 & Write a simple sentence under dictation & M \\
21 & Retrace a complex form & G \\
22 & Copy a telephone number & C \\
23 & Write a telephone number under dictation & M \\
24 & Draw a clock, with all hours and put hands at 11:05 (Clock Drawing Test) & G \\
25 & Copy a paragraph & C \\
\midrule
& \textbf{Subtasks} & \\
26--31 & Subtasks from task 17 & C \\
32--34 & Subtasks from task 14 & M \\
\bottomrule
\end{tabular}
}
\end{table}
\raggedbottom

Certain tasks, such as 14 and 17, involved producing multiple words, which we broke down into subtasks for finer analysis. For instance, task 17 with its six words led to subtasks 26--31, while task 14's three words created subtasks 32--34. This subdivision helped us examine effects like fatigue, where writing might decline more quickly in those with neurodegenerative conditions during repeated efforts.

For data capture, we used a Wacom Bamboo tablet with an ink pen, allowing participants to write naturally on A4 paper. The tablet sampled pen tip positions at 200 Hz, recording coordinates, pressure during contact, and movements when the pen hovered. We segmented each sample into strokes to capture detailed timing aspects of the writing.

From these strokes, we derived a set of features including kinematic and temporal measures, along with participant details, as listed in Table~\ref{tab:featureList}. These features formed the basis for subsequent analysis.

\begin{table}[ht!]
    \caption{Feature list. Feature types are dynamic (D), static (S), and personal (P).} 
    \label{tab:featureList}
    \centering
    \setlength{\tabcolsep}{4pt}
    \begin{adjustbox}{width=\columnwidth}
    \renewcommand{\arraystretch}{1.1}
    \linespread{1}                    
    \rmfamily
    \small                                               
    \begin{tabular}{@{}c l p{10cm} c@{}}
        \toprule
        \textbf{\#} & \textbf{Feature} & \textbf{Description} & \textbf{Type} \\
        \midrule
        1 & Segment  & stroke/Segment number within that trial, for analysis without and with submovement, respectively & S \\
        2 & Start Time & Start time relative to the start of the recording & S \\
        3 & Duration & Time interval between the first and the last points in a stroke & D\\
        4 & Start Vertical Position &	Vertical start position relative to the lower edge of the active digitiser area	&	S	\\
        5 & Vertical Size & Difference between the highest and lowest $y$ coordinates of the stroke	&	S	\\
        6 & Peak Vertical Velocity &	Maximum value of vertical velocity among the points of the stroke	&	D \\
        7 & Peak Vertical Acceleration &	Maximum value of vertical acceleration	among the points of the stroke &	D	\\
        8 & Start Horizontal Position &	Horizontal start position relative to the lower edge of the active tablet area	&	S	\\
        9 & Horizontal Size & Difference between the highest (rightmost) and lowest (leftmost) $x$ coordinates of the stroke	&	S	 \\
        10 & Straightness Error & It is calculated by estimating the length of the straight line, fitting the straight line, estimating the (perpendicular) distances of each point to the fitted line, estimating the standard deviation of the distances, and dividing it by the length of the line between beginning and end &	D \\
        11 & Slant &	Direction from the beginning point to the endpoint of the stroke, in radiant	&	S	\\
        12 & Loop Surface  &	Area of the loop enclosed by the previous and the present stroke	&	S	\\
        13 & Relative Initial Slant & Departure of the direction during the first 80 ms to the slant of the entire stroke.	&	D	\\ 
        14 & Relative Time To Peak Vertical Velocity &	Ratio of the time duration at which the maximum peak velocity occurs (from the start time) to the total duration	&	D \\
        15 & Relative Pen Down Duration & Ratio of pen-up duration to total duration & D \\
        16 & Absolute Size & Calculated from the vertical and horizontal sizes	&	S \\
        17 & Average Absolute Velocity & Average absolute velocity computed across all the samples of the stroke	&	D	\\
        18 & Road length & of a stroke from beginning to end, dimensionless	&	S	\\
        19 & Absolute y Jerk &	The root mean square (RMS) value of the absolute jerk along the vertical direction, across all points of the stroke	&	D	\\
        20 & Normalised y Jerk &	Dimensionless as it is normalised for stroke duration and size	&	D	\\
        21 & Average Normalized y Jerk & Dysfluency measure, theoretically independent of stroke duration and size and marginally dependent upon stroke shape. & D \\
        22 & Absolute Jerk & The Root Mean Square (RMS) value of the absolute jerk across all points of the stroke	&	D	\\
        23 & Normalised Jerk &	Dimensionless as it is normalised for stroke duration and size	&	D	\\
        24 & Average Normalized Jerk & Jerk Average normalised over the strokes & D\\
        25 & Number Of Peak Acceleration Points &	Number of acceleration peaks, both up-going and down-going in the stroke	&	S	 \\
        26 & Average Pen Pressure & Average pen pressure computed over the points of the  stroke	&	D	\\
        27 & Num Of strokes &	Total number of strokes of the task	&	S	\\ 
        \midrule
        32 & Sex	&	Participant's gender	&	P	\\
        33 & Age	&	Participant's age	&	P	\\
        34 & Work	&	Type of work of the participant (intellectual or manual)	&	P	\\
        35 & Education & Participant's education level, expressed in years 	&	P \\
        \bottomrule
    \end{tabular}
\end{adjustbox}
\end{table}

\section{Methods}
\label{sec:methods}

Building on the dataset described, we proceeded with a series of analytical steps to process the data and develop classification models. These methods aimed to extract meaningful patterns from the handwriting dynamics, focusing on temporal aspects that might distinguish AD from healthy controls.

\subsection{Data Preprocessing and Feature Engineering}

To prepare the data for modelling, we applied several preprocessing techniques that accounted for the sequential nature of handwriting.

\subsubsection{Temporal Windowing}

We used a sliding window method to identify temporal patterns in the handwriting sequences. This involved setting window sizes (WS) with stride steps. The overlapping windows helped cover the sequences thoroughly, allowing models to detect both short-term stroke details and longer-term patterns across strokes.

\subsubsection{Feature Normalisation}

Normalisation occurred in stages to manage variations and outliers. First, we applied global scaling with a RobustScaler to kinematic features, which handled extreme values well. Then, we standardised features within each window to adjust for differences in individual writing styles.



\subsection{Deep Learning Architectures}
\label{subsec:dl}

We tested nine neural network designs, each tailored to handle different elements of the temporal data in handwriting.

\subsubsection{Recurrent Neural Networks (RNNs)}

Recurrent Neural Networks form the foundational architecture for sequential data processing, where information flows through a cyclic network structure that maintains memory of previous inputs. The principle underlying RNNs lies in their ability to process sequences of arbitrary length by sharing parameters across time steps, making them particularly suitable for handwriting analysis, where stroke sequences vary considerably in duration and complexity.

Our RNN implementation employs bidirectional processing to capture temporal dependencies from forward and backwards directions within handwriting sequences. This architectural choice addresses the intrinsic limitation of standard RNNs, which can only access past context, by including future information crucial for understanding writing patterns. The bidirectional design enables the model to observe complete stroke trajectories before making classification decisions, which is particularly important when analysing the fluency and coordination patterns that characterise cognitive decline.

\begin{align}
h_t &= \tanh(W_{ih}x_t + W_{hh}h_{t-1} + b_h) \\
\overleftarrow{h_t} &= \tanh(W_{ih}\overleftarrow{x_t} + W_{hh}\overleftarrow{h_{t+1}} + b_h) \\
h_{final} &= [h_T; \overleftarrow{h_1}]
\end{align}

The architecture contains configurable nonlinear activation functions, with the hyperbolic tangent acting as the default choice for its bounded output range and zero-centred properties. Specifically, the hidden state at each time step is computed using the standard RNN update rule (1), and in the backwards direction using (2).  Layer normalisation is applied to stabilise training dynamics, given the variable nature of handwriting kinematics across different subjects and tasks. The final representation combines information from both temporal directions (3), providing a complete encoding of the handwriting sequence that captures progressive deterioration patterns and anticipatory motor planning deficits characteristic of neurodegenerative conditions.

\subsubsection{Long Short-Term Memory Networks (LSTM)}

Long Short-Term Memory networks address the vanishing gradient problem intrinsic in standard RNNs through gating mechanisms that regulate information flow across extended temporal sequences. The LSTM architecture is particularly relevant for handwriting analysis, where subtle motor control deterioration may manifest across varying time scales, from rapid oscillations within individual strokes to gradual changes across writing tasks.

The LSTM's three-gate architecture—forget, input, and output gates—enables selective retention and updating of information throughout the sequence processing. This design addresses the critical challenge in neurodegenerative disease detection, where pathological patterns may emerge gradually and require information integration across extended temporal windows. The forget gate (4) determines which previous information should be discarded, the input gate (5) controls new information, and the output gate (8) regulates which aspects of the cell state (6,7) contribute to the hidden representation (9).

\begin{align}
f_t &= \sigma(W_f \cdot [h_{t-1}, x_t] + b_f) \\
i_t &= \sigma(W_i \cdot [h_{t-1}, x_t] + b_i) \\
\tilde{C_t} &= \tanh(W_C \cdot [h_{t-1}, x_t] + b_C) \\
C_t &= f_t * C_{t-1} + i_t * \tilde{C_t} \\
o_t &= \sigma(W_o \cdot [h_{t-1}, x_t] + b_o) \\
h_t &= o_t * \tanh(C_t)
\end{align}

Our bidirectional LSTM implementation processes handwriting sequences in both temporal directions, capturing dependencies that extend beyond immediate temporal neighbourhoods. This approach is essential for detecting subtle coordination deficits and motor planning impairments characterising early-stage cognitive decline. The architecture contains layer normalisation to stabilise training dynamics and focus on the most discriminative temporal patterns.

\subsubsection{Gated Recurrent Unit (GRU)}

Gated Recurrent Units represent a simplification of the LSTM architecture while maintaining comparable performance for many sequential tasks. The GRU reduces the LSTM's three-gate mechanism into two gates—reset and update—thereby reducing parameter complexity and computational overhead. This architectural efficiency proves particularly advantageous when processing large-scale handwriting datasets where training time and memory constraints become significant factors.

The GRU's design centres on the update gate (11), which determines the balance between retaining previous hidden states and keeping new information, and the reset gate (10), which controls how much of the previous hidden state (13) should influence the candidate activation (12). This dual-gate mechanism enables the model to capture short-term fluctuations in handwriting dynamics and longer-term patterns.

\begin{align}
r_t &= \sigma(W_r \cdot [h_{t-1}, x_t]) \\
z_t &= \sigma(W_z \cdot [h_{t-1}, x_t]) \\
\tilde{h_t} &= \tanh(W \cdot [r_t * h_{t-1}, x_t]) \\
h_t &= (1 - z_t) * h_{t-1} + z_t * \tilde{h_t}
\end{align}

The GRU's reset gate's selective attention to previous states enables the network to focus on relevant historical information while discarding noise, which is particularly important when analysing handwriting from subjects with varying degrees of motor control impairment. The update mechanism promotes smooth transitions between writing phases, capturing the subtle temporal patterns distinguishing healthy motor control from pathological movement signatures.

\subsection{Task Conditioning and Temporal Feature Processing}
\label{sec:task_conditioning}

Handwriting tasks vary significantly in their motor patterns and cognitive demands. To enable the model to adapt to these variations, we condition the neural representations on each specific task using learnable embeddings. This approach supports adaptive processing and promotes knowledge sharing across different tasks.

Each of the 34 tasks is assigned a unique identifier, which is mapped to a 32-dimensional embedding vector via a lookup table. This embedding dimension was chosen to balance representational power with computational cost, scaling appropriately to encode task distinctions without overfitting. The input to our recurrent layers at each time step $t$ is a concatenation of the dynamic stroke features and the corresponding task embedding. 

Formally, for a given task $k$ with identifier $\text{task\_id}_k$, we first obtain its embedding:
\begin{equation}
\mathbf{e}_k = \text{Embedding}(\text{task\_id}_k) \in \mathbb{R}^{32}.
\end{equation}
This task embedding is then concatenated with the 31-dimensional vector of dynamic and kinematic stroke features, $\mathbf{x}_t \in \mathbb{R}^{31}$, for every stroke in the sequence. The resulting vector fed into the models at step $t$ is:
\begin{equation}
\tilde{\mathbf{x}}_t = [\mathbf{x}_t; \mathbf{e}_k] \in \mathbb{R}^{63}.
\end{equation}
This representation allows the recurrent layers to process handwriting dynamics in a task-aware context, with shared layers modelling common motor patterns and the embeddings capturing task-specific adaptations.

\subsection{Integration of Static Subject-Level Features}
\label{sec:static_features}

Static subject-level data, such as demographic information (age, sex, education), do not vary over the course of a handwriting task. Feeding such constant information into a recurrent layer at every time step is inefficient and architecturally mismatched, as recurrent models are designed to model temporal change. 

To address this, we adopt a hybrid architecture where static features are integrated only after the temporal processing is complete. First, the models processes the entire sequence of task-conditioned stroke vectors ($\tilde{\mathbf{x}}_1, \dots, \tilde{\mathbf{x}}_T$) and produces a final hidden state vector, $\mathbf{h}_{\text{final}}$, which serves as a summary of the sequence's temporal dynamics. This vector is then concatenated with the subject's 4-dimensional static feature vector, $\mathbf{p}_s \in \mathbb{R}^{4}$.

The final combined feature vector, $\mathbf{z}_s$, is defined as:
\begin{equation}
\mathbf{z}_s = [\mathbf{h}_{\text{final}}; \mathbf{p}_s].
\end{equation}
This vector, which now encodes both the learned temporal patterns from the handwriting task and the static context of the individual, is passed to a final feed-forward classification head to produce the probability of an AD. This architecture ensures that temporal and static features are processed by the most appropriate model components, respecting their distinct natures.

\subsection{Training Strategy and Optimisation}

The training strategy uses a robust cross-validation framework, carefully designed loss functions, and optimisation techniques to ensure reliable performance assessment and effective model convergence.

\subsubsection{Cross-Validation Framework}

The experimental design employs a 5-fold stratified cross-validation methodology adapted for subject-level evaluation, addressing the challenge of maintaining subject independence across training and testing phases. This approach prevents data leakage that could occur if samples from the same subject appeared in both training and testing sets, ensuring that performance metrics reflect real generalisation capabilities to unseen individuals rather than memorisation of subject-specific patterns.

The stratification process maintains balanced representation of healthy controls and AD patients across all folds, with each fold preserving the original class distribution. This balanced partitioning is critical given the task's clinical nature, where sensitivity and specificity must be optimised simultaneously. The subject-level stratification operates by grouping all handwriting samples by subject identifier, then distributing subjects across folds while maintaining class balance.

For each fold iteration, a distinct random seed is generated as $seed_{fold} = seed_{base} + fold_{index}$, ensuring reproducible yet varied initialisations across folds. This deterministic randomisation enables exact replication of experimental results while maintaining statistical independence between fold evaluations. The cross-validation independently processes each WS and stride configuration, generating complete performance profiles across different temporal granularities.

The validation strategy incorporates early stopping mechanisms based on F1 score monitoring in holdout validation sets within each fold. This approach prevents overfitting while optimising for balanced classification performance, which is particularly important in medical applications where both false positives and false negatives carry significant clinical implications. The validation frequency is set to half-epoch intervals, providing fine-grained monitoring of model convergence without excessive computational overhead.

\subsubsection{Loss Function}

The classification objective employs binary cross-entropy loss with optional class weighting to address potential class imbalance in the clinical dataset. The loss function is formulated as (16):

\begin{align}
\mathcal{L} = -\frac{1}{N} \sum_{i=1}^{N} w_{y_i} \left[ y_i \log(\hat{y}_i) + (1-y_i) \log(1-\hat{y}_i) \right]
\end{align}

where $N$ represents the batch size, $y_i \in \{0,1\}$ denotes the true label for sample $i$ (0 for healthy controls, 1 for AD), $\hat{y}_i$ represents the predicted probability, and $w_{y_i}$ denotes the class-specific weight.

The class weights are computed using inverse frequency weighting to mitigate the effects of class imbalance (17):

\begin{align}
w_c = \frac{N}{2 \cdot N_c}
\end{align}

where $N_c$ represents the number of samples in class $c$, and $N$ is the total number of samples. This weighting scheme ensures that the minority class receives adequate emphasis during training, promoting balanced learning.

\subsubsection{Optimization}

The optimisation strategy employs the AdamW optimiser. This optimiser utilises adaptive moment estimation with bias correction and improved weight decay handling (18):

\begin{align}
\theta_{t+1} &= \theta_t - \eta \left( \frac{\hat{m}_t}{\sqrt{\hat{v}_t} + \epsilon} + \lambda \theta_t \right)
\end{align}

where $\theta_t$ represents the model parameters at iteration $t$, $\eta$ denotes the learning rate, $\hat{m}_t$ and $\hat{v}_t$ are bias-corrected first and second moment estimates, $\epsilon$ is a small constant for numerical stability, and $\lambda$ represents the weight decay coefficient.

The learning rate is initially set to $1 \times 10^{-4}$ with adaptive scheduling through \emph{ReduceLROnPlateau}, which monitors validation F1-score and reduces the learning rate by a factor of 0.5 when performance plateaus persist for more than five epochs. The minimum learning rate is bounded at $1 \times 10^{-6}$ to prevent excessive reduction that could halt learning progress.

Weight decay regularisation is set to $5 \times 10^{-5}$, providing moderate regularisation to prevent overfitting while maintaining model expressiveness. 

Gradient clipping is implemented with a maximum norm constraint of 1.0 to prevent gradient explosion, a common challenge in recurrent network training. The clipping operation normalises gradients when their norm exceeds the threshold:

\begin{align}
\nabla\theta \leftarrow \begin{cases}
\nabla\theta & \text{if } \|\nabla\theta\| \leq \text{clip\_val} \\
\frac{\text{clip\_val}}{\|\nabla\theta\|} \nabla\theta & \text{otherwise}
\end{cases}
\end{align}

The training process incorporates mixed precision computation utilising 16-bit floating point for forward passes and 32-bit precision for gradient computations. This approach reduces memory consumption and accelerates training.

The framework includes monitoring gradient statistics through custom callbacks that track gradient norms, parameter updates, and learning rate evolution. These metrics provide insights into training dynamics and enable early detection to avoid vanishing or exploding gradients.

The deterministic training configuration ensures reproducible results across multiple runs, enabling precise replication of experimental findings.

\subsection{Experimental Setup}
\label{subsec:expr}

We conducted a preliminary exploratory data analysis before model development to identify optimal segmentation parameters. This step focused on determining window sizes ($w$) and stride values ($s$) that would preserve meaningful temporal patterns while minimising redundancy in the sequential data. To guide this process, we formulated a heuristic metric, the \textit{Temporal Stability Score (TSS)}, designed to quantify the suitability of different segmentation configurations. This score provides a principled way to balance the trade-off between information preservation and computational efficiency. 

Formally, for each window size $w$ and stride $s$, we define the TSS as:
\[
\mathrm{TSS}(s, w) = D_s(w) + A(w) - R(s) + E(w)
\]
In this formulation, we assign equal importance to each positive component by setting their implicit weights to unity and explicitly penalise redundancy. The components are defined as follows:
\begin{itemize}
    \item $D_s(w)$ quantifies \textit{stroke count stability}, based on the distributional consistency of stroke counts across subjects and tasks.
    \item $A(w)$ measures \textit{autocorrelation persistence}, defined as the lag at which the autocorrelation function decays below a threshold (0.2), capturing within-stroke temporal dependencies.
    \item $R(s)$ reflects \textit{stride-induced redundancy}, which is subtracted to penalise high overlap between consecutive windows, estimated via normalised mutual information.
    \item $E(w)$ denotes the \textit{variance-normalised entropy}, capturing intra-window signal complexity while compensating for amplitude variability across subjects.
\end{itemize}
By optimising for this score, we aimed to select parameters that robustly represent the underlying dynamics of handwriting for subsequent model training.

Optimal parameters $(s^*, w^*)$ are selected by maximising $\mathrm{TSS}(s, w)$, achieving a principled balance between information preservation and redundancy minimisation across temporal scales.

Based on these findings, we implemented the three architectures mentioned in Section~\ref{subsec:dl} to evaluate their temporal pattern recognition in the segmented data. Models were created using PyTorch Lightning and trained on NVIDIA GPUs with CUDA acceleration. The training configuration employed batch sizes of 64 samples, hidden layer dimensions between 128 and 256 units, dropout regularisation at 0.3, and layer normalisation to stabilise gradient flow.
We evaluated each architecture across nine configurations, testing window sizes of 60, 70, and 80 samples with strides of 1, 2, and 5. This evaluation allowed us to assess the impact of architectural choices and the sensitivity of each model to temporal segmentation parameters. 
To provide a robust performance benchmark, we compared our results against several traditional ML ensemble methods. The methodology for these is drawn from our prior work~\cite{nardone2025handwriting}, and we replicated the experiments on the identical dataset and cross-validation folds to ensure a direct and fair comparison.
A crucial distinction in this approach is the input representation. Unlike the sequential windows processed by the recurrent models, these traditional ML models operate on feature vectors from \textit{individual strokes}. Each stroke is treated as an independent data point, and its features are used to train base classifiers. The temporal relationship between consecutive strokes is therefore not explicitly modelled. Four different ensemble strategies were then employed to aggregate these stroke-level predictions into a final, subject-level classification: majority voting (MV), weighted majority voting (WMV), ranking-based aggregation, and stacking. By using the same subject-level stratified evaluation protocol, we can directly compare the generalisation capabilities of these models against the sequence-aware recurrent architectures.

\subsection{Evaluation Metrics}

The performance of the model was assessed through three complementary metrics that provide an exhaustive insight into the classification capacity. Accuracy measures the overall proportion of correctly classified samples across all classes. Sensitivity quantified the model's ability to correctly identify positive cases, thus AD patients, while specificity evaluated performance on negative cases, healthy controls. These metrics were computed using 5-fold cross-validation to ensure robust performance estimates, with mean and standard deviation reported across folds to capture model stability and generalisation capacity.

\section{Results}
\label{sec:results}

This section reports the experimental findings obtained throughout the study. The results are organised following experiments described in Section~\ref{subsec:expr}, starting from hyperparameter finding, model performance across various recurrent neural architectures, and comparisons with traditional machine learning baselines.

\begin{figure}[t]
    \centering
    \includegraphics[width=\columnwidth, clip, trim=4cm 15.1cm 3.5cm 5cm]{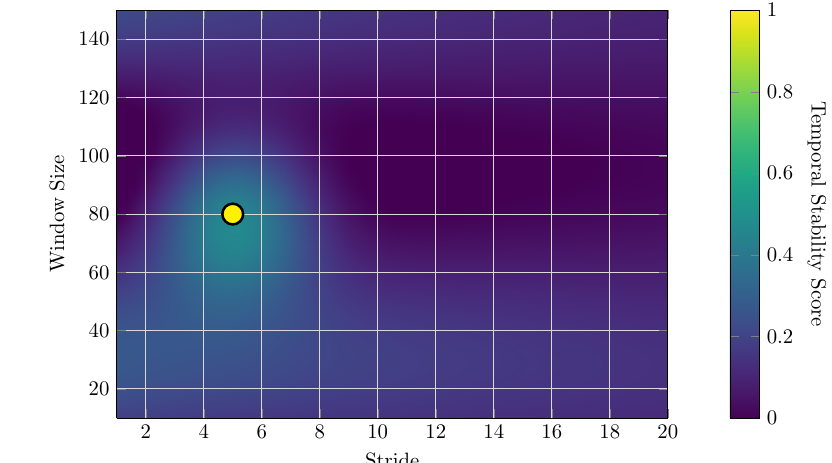}
    \caption{Surface plot representing the temporal stability score across
different window sizes and strides.}
    \label{fig:eda_hyperparameter_surface}
\end{figure}

The first experiment focused on identifying suitable segmentation parameters for temporal sequence modelling. The analysis was conducted prior to model training to inform hyperparameter selection through a data-driven approach.
We evaluated a range of window sizes and stride values to assess their impact on temporal structure preservation within the dataset. Specifically, we computed the \textit{temporal stability score} over a grid of configurations, resulting in a surface representation that reflects the statistical consistency of temporal segments.

\begin{table}[t]
\centering
\caption{Average test performance metrics in percentages (mean (standard deviation)) for the RNN model.}
\label{tab:rnn_test_metrics}
\adjustbox{width=0.5\textwidth,center}{
\begin{tabular}{@{}llccc@{}}
\toprule
\multirow{2}{*}{WS} & \multirow{2}{*}{Stride} & \multirow{2}{*}{Accuracy} & \multirow{2}{*}{Sensitivity} & \multirow{2}{*}{Specificity} \\
& & & & \\
\midrule
\multirow{3}{*}{60} & 1 & 61.8 (26.3) & 45.3 (20.1) & 72.8 (33.0) \\
& 2 & 49.6 (26.2) & 23.1 (11.9) & 77.9 (34.4) \\
& 5 & 59.7 (19.0) & 83.3 (37.4) & 39.1 (41.4) \\
\multirow{3}{*}{70} & 1 & 55.8 (31.9) & 58.3 (31.2) & 60.0 (49.6) \\
& 2 & 49.7 (25.2) & 33.3 (17.1) & 63.3 (42.4) \\
& 5 & 46.5 (25.5) & 26.1 (37.4) & 62.6 (42.5) \\
\multirow{3}{*}{80} & 1 & 66.2 (13.5) & 35.6 (18.8) & 81.9 (20.8) \\
& 2 & 60.9 (28.6) & 49.0 (35.7) & 68.5 (46.3) \\
& 5 & 67.9 (24.1) & 62.5 (42.6) & 20.0 (44.7) \\
\bottomrule
\end{tabular}
}
\end{table}

Figure~\ref{fig:eda_hyperparameter_surface} shows the resulting surface, which highlights a region of high temporal stability corresponding to window sizes between 60 and 80 samples and stride values between 1 and 5. Among these, the configuration with a stride of 5 and a window size of 80 emerged as the most promising. This combination achieves a principled balance between dense temporal coverage and computational efficiency, while preserving the intrinsic dynamics of the signal. These findings informed the segmentation strategy for subsequent model development using recurrent neural architectures.

Table \ref{tab:rnn_test_metrics} shows that RNN exhibits significant performance variability across configurations, with standard deviations often exceeding 20 percentage points. The model reveals a trade-off between sensitivity and specificity, particularly in the WS=80, Stride=5 configuration, which achieves the highest accuracy (67.9\%) but suffers from extremely poor specificity (20.0\%). This imbalance suggests the model has learned to favour positive class predictions.

From Table \ref{tab:lstm_test_metrics}, the LSTM consistently achieves high sensitivity across most configurations (often exceeding 70\%), but at the cost of specificity, which rarely surpasses 50\%. The model performs more stably than the RNN, with generally lower standard deviations. The persistent sensitivity-specificity imbalance across all WS and stride combinations indicates this is likely an inherent limitation of the LSTM's learning dynamics on this dataset rather than a hyperparameter-dependent phenomenon.

\begin{table}[b]
\centering
\caption{Average test performance metrics in percentages (mean (standard deviation)) for the LSTM model.}
\label{tab:lstm_test_metrics}
\adjustbox{width=0.5\textwidth,center}{
\begin{tabular}{@{}llccc@{}}
\toprule
\multirow{2}{*}{WS} & \multirow{2}{*}{Stride} & \multirow{2}{*}{Accuracy} & \multirow{2}{*}{Sensitivity} & \multirow{2}{*}{Specificity} \\
& & & & \\
\midrule
\multirow{3}{*}{60} & 1 & 61.1 (20.7) & 83.4 (21.3) & 46.5 (37.2) \\
& 2 & 54.7 (22.8) & 89.2 (19.2) & 30.1 (37.6) \\
& 5 & 62.3 (18.7) & 63.6 (37.6) & 62.0 (39.8) \\
\multirow{3}{*}{70} & 1 & 57.1 (25.8) & 76.1 (16.1) & 42.6 (37.6) \\
& 2 & 54.7 (23.8) & 73.7 (18.1) & 35.1 (39.1) \\
& 5 & 54.1 (23.4) & 79.4 (29.1) & 34.6 (38.9) \\
\multirow{3}{*}{80} & 1 & 61.8 (31.6) & 72.3 (20.9) & 51.8 (42.3) \\
& 2 & 61.8 (31.6) & 74.1 (15.9) & 43.4 (42.4) \\
& 5 & 64.5 (18.3) & 78.7 (25.0) & 32.2 (35.6) \\
\bottomrule
\end{tabular}
}
\end{table}

Table \ref{tab:gru_test_metrics} (GRU Performance) reveals that GRU has the most extreme sensitivity bias among the three architectures, achieving remarkable sensitivity scores (up to 90.1\%) while maintaining consistently poor specificity. The WS=60, Stride=1 configuration shows the lowest variance in sensitivity (7.5\%) among all DL experiments, suggesting stable but heavily biased predictions. This pattern indicates the GRU's gating mechanism may be particularly susceptible to overfitting to the positive class in this temporal classification task.

Among the DL models GRU (WS=60, stride=1), obtains the highest sensitivity, making it effective at identifying AD cases but with a lower specificity, suggesting a tendency to misclassify healthy controls; LSTM (WS=80, stride=5) offers a more balanced profile, pointing to its robustness in integrating long-term temporal dependencies; while RNN (WS=80, stride=5), although simpler, performs competitively, showing potential when long stroke sequences are considered. However, standard deviations are relatively high, reflecting instability possibly due to the limited dataset size and variability among handwriting tasks.

\begin{table}[t]
\centering
\caption{Average test performance metrics in percentages (mean (standard deviation)) for the GRU model.}
\label{tab:gru_test_metrics}
\adjustbox{width=0.5\textwidth,center}{
\begin{tabular}{@{}llccc@{}}
\toprule
\multirow{2}{*}{WS} & \multirow{2}{*}{Stride} & \multirow{2}{*}{Accuracy} & \multirow{2}{*}{Sensitivity} & \multirow{2}{*}{Specificity} \\
& & & & \\
\midrule
\multirow{3}{*}{60} & 1 & 62.4 (17.5) & 90.1 (7.5) & 40.4 (34.4) \\
& 2 & 61.8 (18.9) & 87.0 (14.3) & 37.6 (36.4) \\
& 5 & 55.1 (16.5) & 75.0 (37.4) & 34.7 (34.6) \\
\multirow{3}{*}{70} & 1 & 57.3 (18.3) & 89.1 (14.3) & 31.6 (30.5) \\
& 2 & 61.4 (16.5) & 88.5 (14.7) & 34.4 (35.3) \\
& 5 & 56.7 (19.6) & 75.0 (37.4) & 40.4 (34.6) \\
\multirow{3}{*}{80} & 1 & 60.0 (25.0) & 71.4 (18.9) & 49.7 (42.3) \\
& 2 & 51.8 (29.8) & 78.3 (19.9) & 32.2 (35.6) \\
& 5 & 54.0 (21.0) & 78.7 (25.0) & 30.4 (35.6) \\
\bottomrule
\end{tabular}
}
\end{table}

Finally, in Table \ref{tab:dl_al_full_comparison}, the comparison reveals a performance gap between DL and ensemble ML approaches. While the DL models exhibit high variance and severe class prediction imbalance, the ML methods achieve higher accuracy and remarkably balanced sensitivity-specificity trade-offs. The ranking-based ensemble achieves the best overall performance, particularly the Ranking-based ensemble (80.18\% accuracy), with nearly symmetric sensitivity and specificity values, suggesting these methods better capture the underlying data structure without overfitting to class distributions.

Finally, we compare the performance of the recurrent architectures against the traditional ML ensemble baselines, whose methodology is detailed in Section~\ref{sec:methods}. The results, presented in Table~\ref{tab:dl_al_full_comparison}, reveal a performance gap. While the deep learning models consistently suffer from high variance and severe class prediction imbalance, the traditional ML methods achieve substantially higher accuracy and remarkably balanced sensitivity-specificity trade-offs.

The ranking-based ensemble, in particular, yields the best overall performance (80.18\% accuracy) with nearly symmetric sensitivity and specificity. This outcome strongly supports our central thesis: the architectural assumptions of recurrent models are fundamentally mismatched with pre-segmented, stroke-level feature data. By treating each stroke as an independent sample, the traditional models avoid the pitfalls of processing artificially constrained sequences and prove far more effective at capturing the discriminative patterns for this diagnostic task.

\begin{table}[b]
\centering
\caption{Comparison of DL models and ML ensembles. Values are shown as percentage mean (standard deviation).}
\label{tab:dl_al_full_comparison}
\begin{adjustbox}{width=0.75\textwidth}
\begin{tabular}{lccc}
\toprule
\textbf{Method} & \textbf{Accuracy (\%)} & \textbf{Sensitivity (\%)} & \textbf{Specificity (\%)} \\
\midrule
RNN (WS=80, Stride=5) & 67.90 (24.10) & 62.50 (42.60) & 20.00 (44.70) \\
LSTM (WS=80, Stride=5) & 64.50 (18.30) & 78.70 (25.00) & 32.20 (35.60) \\
GRU (WS=60, Stride=1) & 62.40 (17.50) & 90.10 (7.50) & 40.40 (34.40) \\
\midrule
MV (ML) & 77.90 (5.26) & 78.55 (5.45) & 77.25 (5.38) \\
WMV (ML) & 74.81 (9.15) & 75.48 (9.35) & 74.15 (9.28) \\
Ranking (ML) & \textbf{80.18 (6.22)} & \textbf{80.85 (6.42)} & \textbf{79.52 (6.35)} \\
Stacking (ML) & 77.30 (9.50) & 77.95 (9.68) & 76.65 (9.62) \\
\bottomrule
\end{tabular}
\end{adjustbox}
\end{table}

\section{Conclusions}
\label{sec:conclusions}
The proposed study explored deep models for distinguishing AD patients from healthy controls based on handwriting dynamics. While DL shows its potential in temporal sequence modelling, our findings highlight some key limitations that limit its performance in this case study.
First, recurrent models do not perform effectively on stroke-level inputs. A critical issue comes from the lack of a universally accepted definition of a "stroke", as it depends on multiple segmentation criteria, depending on pen-up/pen-down events and velocity inversions. This ambiguity breaks the natural continuity that RNNs are designed to exploit. As a result, the temporal dependencies they model may be misaligned with the underlying motor sequences of handwriting.

Second, the models were trained on pre-extracted features from already segmented strokes, not on continuous, raw time-series data. This detachment from raw temporal signals limits the models' ability to learn meaningful dynamics, undermining the potential of recurrence-based architectures. In this way, we are limiting the power of the models that operate on artificially constrained sequences, reducing their ability to capture true neuromotor deterioration. 

Third, although task embeddings were used to mitigate inter-task variability, the performance of deep models still fluctuated depending on the task type, as shown by the metrics' standard deviation. This suggests the need for more advanced mechanisms for modelling task context, or a shift toward subject-level integration, where models can learn from richer, aggregated representations.

Although the study has limitations and has not yet achieved strong predictive performance with deep models, it nevertheless contributes by highlighting several critical aspects that deserve further investigation. It also underscores and suggests future directions for improving the temporal representation of handwriting data and the design of learning architectures.

The first thing to take into account is to move to raw time-series inputs, enabling models to directly learn temporal patterns without relying on heuristic stroke segmentation. Another direction that can be considered is the exploration of different models, which may better accommodate irregular, non-stationary sequences. Finally, the work can be improved by investigating more on task-aware or subject-aware learning strategies, such as attention-based fusion or hierarchical modelling, to capture both task-level variability and subject-specific patterns.
Ultimately, while DL remains a promising approach, its full potential can only be realised by aligning model design with the true structure and granularity of the data, moving beyond stroke-level abstractions to capture the full richness of handwriting as a cognitive-motor process.





\section*{Hardware and Infrastructure}

\begin{figure}[t]
    \centering
    \includegraphics[width=\columnwidth, clip, trim=4.8cm 5cm 4.8cm 5cm]{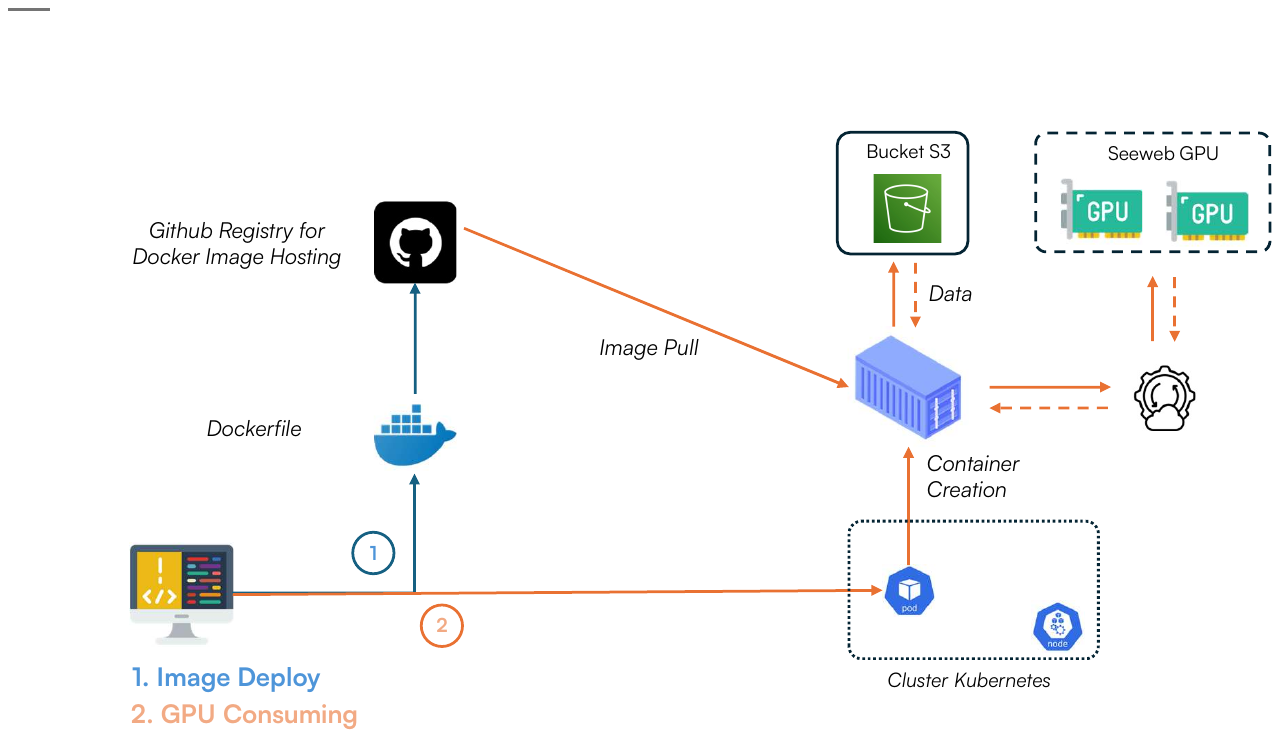}
    \caption{Serverless architecture for GPU consuming.}
    \label{fig:kuberscheme}
\end{figure}

All DL experiments were performed on a high-performance computing infrastructure provided by \textit{Seeweb s.r.l.},  using their dedicated NVIDIA RTX\texttrademark{} A6000 GPUs. Our experimental methodology was architected for reproducibility and scalability using a container-based workflow.

The architecture, shown in Figure~\ref{fig:kuberscheme}, is centred around containerisation and orchestration. Application environments were defined using Dockerfiles, and the resulting Docker images were stored and versioned in the GitHub Container Registry. A Kubernetes cluster handled these containerised workloads' deployment, scaling, and management. Upon initiating an experimental run, a Kubernetes deployment would schedule a pod on a GPU-enabled node. The node would then pull the requisite container image from the registry. The running container performed its computational tasks by accessing datasets from an S3-compatible storage bucket and executing deep learning models on the allocated NVIDIA GPU resources. This serverless-style consumption of GPU resources provided an efficient and on-demand environment for our research.

\section*{Data Availability Statement}

The datasets analysed in this study are available from the corresponding author upon reasonable request, subject to privacy and ethical considerations.




\footnotesize
\section*{Acknowledgements}
The authors gratefully acknowledge \textit{Seeweb s.r.l.} \cite{seeweb_home} for providing computational resources and access to their Cloud Server GPU infrastructure \cite{seeweb_gpu} used to perform deep learning experiments.


\end{document}